\documentclass[aps,prb,showpacs,twocolumn,floats]{revtex4}
\usepackage{graphicx}
\usepackage{subfigure}
\usepackage{epsfig}
\usepackage{dcolumn}
\usepackage{bm}
\usepackage[ansinew]{inputenc}
\usepackage{amsmath}
\usepackage{amsthm}
\usepackage[T1]{fontenc}
\usepackage{amssymb}
\usepackage{amsfonts}
\usepackage[english]{babel}
\usepackage{enumitem}

\newcommand{\ud}{\,\mathrm{d}}
\newcommand{\ds}{\displaystyle}

\begin{document}

\title{Excitation modes of vortices in sub-micron magnetic disks}
\date{\today}
\author{R. Zarzuela$^{1,2}$, E. M. Chudnovsky$^{2}$, J. Tejada$^{1}$}
\affiliation{$^{1}$Departament de F\'{i}sica Fonamental, Facultat
de F\'{i}sica, Universitat de Barcelona, Avinguda Diagonal 645,
08028 Barcelona, Spain\\ $^{2}$Physics Department, Lehman College,
The City University of New York, 250 Bedford Park Boulevard West,
Bronx, NY 10468-1589, U.S.A.}
\date{\today}

\begin{abstract}
Classical and quantum theory of spin waves in the vortex state of
a mesoscopic sub-micron magnetic disk has been developed with
account of the finite mass density of the vortex. Oscillations of
the vortex core resemble oscillations of a charged string in a
potential well in the presence of the magnetic field. Conventional
gyroscopic frequency appears as a gap in the spectrum of spin
waves of the vortex. The mass of the vortex has been computed that
agrees with experimental findings. Finite vortex mass generates a
high-frequency branch of spin waves. Effects of the external
magnetic field and dissipation have been addressed.
\end{abstract}


\maketitle

\section{Introduction}

Recent advances in optical and electron-beam lithography offered
possibility to fabricate arrays of micron and submicron-size
magnetic structures with controlled magnetic properties. Among
such structures are mesoscopic circular disks of soft
ferromagnetic materials. Arrays of such disks, as well as
individual disks, have been intensively studied
\cite{Cowburn,Shinjo,Novosad1,Novosad2,GdL,Castel,Zarzuela} due to
their unusual magnetic properties and potential for technological
and biomedical applications\cite{Parkin,Rozhkova,Kim}.

Micron-size circular disks exhibit a wide variety of magnetic
equilibrium configurations due to geometric constraints on the
spin field\cite{Hertel}. Their applications are based on static
and dynamic properties of one of the essentially non-uniform
ground states, the \emph{vortex state}. It is characterized by the
curling of the magnetization in the plane of the disk, leaving
virtually no magnetic ``charges''. The very weak uncompensated
magnetic moment of the disk sticks out of a small area confined to
the vortex core (VC). The diameter of the core is comparable to
the material exchange length\cite{Novosad1,Guslienko1}. The low
frequency dynamics of the vortex state is due to the gyrotropic
mode, consisting of the spiral-like precessional motion of the VC
as a whole\cite{Choe,Guslienko2,Guslienko3,Guslienko4,Lee}, and it
is intrinsically distinct from conventional spin wave excitations.

Because of the strong exchange interaction among the out-of-plane
spins in the VC, it behaves as an independent entity. The research
on excitation modes of vortices has focused on the low-frequency
gyroscopic mode that describes circular motion of the vortex about
the center of the disk. It can also be viewed as the uniform
precession of the magnetic moment of the disk due to the vortex.
The natural question is whether the gyroscopic mode allows spatial
dispersion similar to spin waves of finite wavelength in
ferromagnets. The aim of this paper is to study spin waves related
to the gyroscopic motion of the vortex. Such a wave is shown in
Fig. 1. It must exist due to finite elasticity of the vortex
provided by the exchange interaction.

\begin{figure}[htbp!]
\center
\includegraphics[scale=0.3]{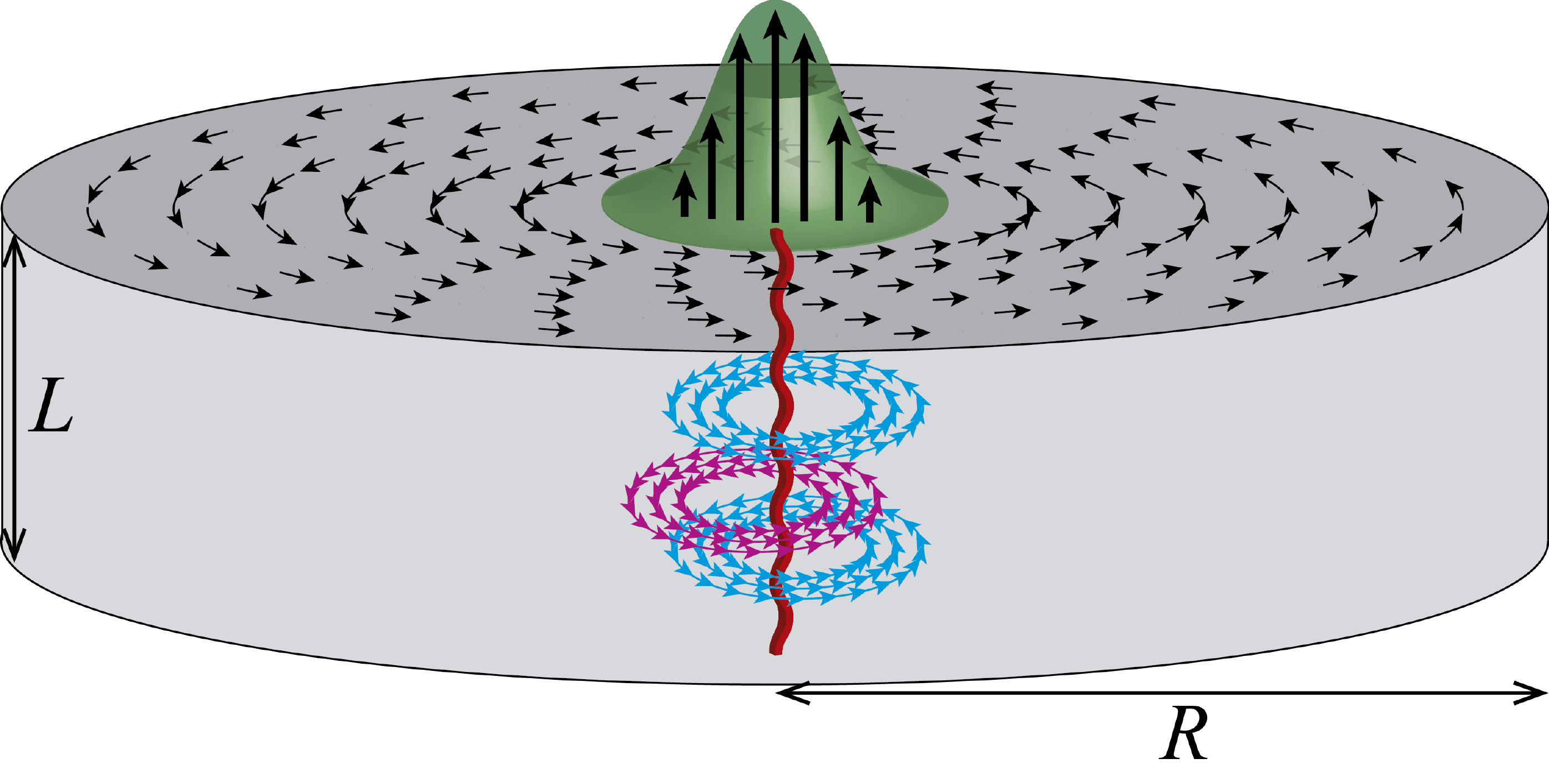}
\caption{Gyroscopic spin wave in the vortex state of a mesoscopic
magnetic disk.}
\end{figure}

Most of the research on the gyroscopic motion of vortices in
circularly polarized disks ignores the inertial mass of the VC.
Such a mass has a dynamical origin stemming from the variation of
the shape of the VC as it moves inside the disk. Meantime,
experimental studies of vortex oscillations in micrometer
permalloy rings\cite{Bedau} hinted towards a non-negligible vortex
mass of order $10^{-24}$kg. On a theoretical side the vortex mass
has been previously computed in a two-dimensional Heisenberg model
with anisotropic exchange interaction \cite{Gouvea,Wysin,Ivanov}.
In disks made of soft magnetic materials that have been
experimented with, the exchange interaction is isotropic. We will
show that in this case the finite mass density of the vortex
originates from the geometrical confinement of the spin field and
the magnetic dipole-dipole interactions.

In this paper we will derive the generalized Thiele equation that
describes spin waves in the vortex core of finite mass density and
will obtain the spectrum of such waves. We will show that the
conventional gyroscopic mode, $\omega_G$, appears as a gap in the
spectrum of the spin waves in the vortex, $\omega(q) = \omega_{G}
+ \alpha q^2$, when the vortex mass is neglected. From the
mathematical point of view the above problem resembles the problem
of the motion of a charged string in a potential well in the
presence of the magnetic field. The latter problem is a
generalization of the problem of Landau levels of an electron in a
two-dimensional potential well in the magnetic field. We will show
that this problem has a nice exact solution for quantized
oscillations of the string, thus providing the spectrum of magnons
in the vortex in the quantum regime as well. Classical and quantum
solutions for the spectrum of excitations of the vortex lead to
the same dispersion law, $\omega(q)$, in the limit of small $q$.

The paper is structured as follows. In Sec. II a Lagrangian
formulation of the problem is presented. Formal derivation of the
massive elastic Thiele's equation that allows deformations of the
vortex line is given in Section III. The spectrum of spin waves in
the vortex core is obtained in Section IV. We show that a finite
mass of the vortex results in the additional excitation mode that
is absent in the case of zero mass. Quantum mechanical treatment
of magnons in the vortex core is developed in Section V. The
vortex mass in a circularly polarized disk is computed in Sec. VI
and is shown to be in good agreement with experimental findings.
The field dependence of the vortex excitation modes and effects of
dissipation are discussed in Sec. VII. Sec. VIII contains final
conclusions and suggestions for experiment.

\section{Lagrangian mechanics of the vortex core}
We shall describe the vortex line by the vector field
$\vec{X}=(x,y)$, where $x(t,z)$ and $y(t,z)$ are coordinates of
the center of the vortex core in the $XY$ plane. Landau-Lifshitz
dynamics of the fixed-length magnetization vector
$\vec{M}(\Theta,\Phi)=M_{s}(\cos\Phi\sin\Theta,\sin\Phi\sin\Theta,\cos\Theta)$
follows from the Lagrangian \cite{CT-lectures}
\begin{align}
\label{Lagrangian}
 \mathcal{L}&\Big[t;\Theta,\Phi,\dot{\Theta},\dot{\Phi},\partial_{z}\Theta,\partial_{z}\Phi\Big]
 =\nonumber\\
 &\int\ud z\ud^{2}\vec{r}\Bigg[\frac{M_{s}}{\gamma}(D_{t}\Phi)\cos\Theta-\mathcal{E}(\Theta,\Phi, \partial_{z}\Theta,\partial_{z}\Phi)\Bigg]
\end{align}
where
$\mathcal{E}(\Theta,\Phi,\partial_{z}\Theta,\partial_{z}\Phi)$ is
the energy density. The dependence of the Lagrangian on the
partial derivatives $\partial_{z}$ of the angular coordinates
comes from the elastic nature of the vortex core. It is contained
in the total energy, $\mathcal{E}$, that takes into account
interaction between different layers of the vortex line, see
below.

The spatial dependence of angular coordinates $(\Theta, \Phi)$ for
the vortex state is given by
$\Theta=\Theta(t;\vec{r},z)=\Theta(\vec{r}-\vec{X}(t,z),t)$ and
$\Phi=\Phi(t;\vec{r},z)=\Phi(\vec{r}-\vec{X}(t,z),t)$. We only
consider long-wave solutions that do not deform the vortex core in
any $z$-cross-section of the disk. This means that the angular
coordinates depend on $t$ and $z$ via the coordinates of the
vortex core $\vec{X}(t,z)$. The covariant derivative with respect
to time, $D_{t}\Phi$, along the vortex core is given by
\begin{equation}
\ds
D_{t}\Phi=\nabla_{\dot{\vec{X}}(t,z)}\Phi=-\dot{\vec{X}}(t,z)\cdot\nabla_{\vec{r}}
\Phi(\vec{r}-\vec{X}(t,z))
\end{equation}
where ``dot'' denotes partial derivative with respect to $t$.

Taking all these considerations into account, the above Lagrangian
becomes
\begin{equation}
\label{Lagrangian2}
 \mathcal{L}\left[t;\vec{X},\dot{\vec{X}},\partial_{z}\vec{X}\right]=\int\ud z\tilde{\mathcal{L}}
 \left[t,z;\vec{X},\dot{\vec{X}},\partial_{z}\vec{X}\right]
 \end{equation}
 with the Lagrangian density being
 \begin{align}
 \label{Lagdensity}
\tilde{\mathcal{L}}\left[t,z;\vec{X},\dot{\vec{X}},\partial_{z}\vec{X}\right]&=\int\ud^{2}\vec{r}
\Bigg[\frac{M_{s}}{\gamma}\left(-\dot{\vec{X}}(t,z)\cdot\nabla_{\vec{r}}\Phi\right)\cos\Theta\nonumber\\
&-\mathcal{E}(\Theta,\Phi,\partial_{z}\Theta,\partial_{z}\Phi)\Bigg].
\end{align}
Thus the generalized momentum densities are given by
\begin{eqnarray}
 \label{momentums}
  \vec{\Pi}_{t}\left[t,z;\vec{X},\dot{\vec{X}},\partial_{z}\vec{X}\right]&\equiv&
  \frac{\delta\tilde{\mathcal{L}}}{\delta\left(\dot{\vec{X}}(t,z)\right)}\nonumber\\
  &=& -\frac{M_{s}}{\gamma}\int\ud^{2}\vec{r}\left(\nabla_{\vec{r}}\Phi\right)\cos\Theta\\
  \vec{\Pi}_{z}\left[t,z;\vec{X},\dot{\vec{X}},\partial_{z}\vec{X}\right]
  &\equiv&\frac{\delta\tilde{\mathcal{L}}}{\delta\left(\partial_{z}\vec{X}(t,z)\right)}\nonumber\\
  &=&-\frac{\delta\omega(\vec{X},\partial_{z}\vec{X})}{\delta(\partial_{z}\vec{X}(t,z))}
\end{eqnarray}
with $\ds
\omega(\vec{X},\partial_{z}\vec{X})=\int\ud^2\vec{r}\;\mathcal{E}(\Theta,\Phi,\partial_{z}\Theta,\partial_{z}\Phi)$
being the linear energy density. The dynamics of the vortex core
is governed by the Euler-Lagrange equation,
\begin{equation}
\ds
D_{t}\vec{\Pi}_{t}+\partial_{z}\vec{\Pi}_{z}-\frac{\delta\tilde{\mathcal{L}}}
{\delta\vec{X}(t,z)}=0
\end{equation}
Notice that
\begin{align}
D_{t}\Lambda(t,z;\;&\vec{r},\vec{v})=\dot{\xi}(t,z)(\Lambda)=\frac{\partial}{\partial
t}\Lambda(\vec{r}-\vec{X}(t,z),\dot{\vec{X}}(t,z)) \nonumber \\
&=-\dot{\vec{X}}(t,z)\cdot\nabla_{\vec{r}}\Lambda(\vec{r}-\vec{X}(t,z),\dot{\vec{X}}(t,z))
+\nonumber\\
&\qquad\ddot{\vec{X}}(t,z)\cdot\nabla_{\vec{v}}\Lambda(\vec{r}-\vec{X}(t,z),\dot{\vec{X}}(t,z))
\end{align}
means covariant derivative along the curve that is tangent to the
vortex core, $\xi(t,z)$. All terms involving $\dot{\vec{X}}(t,z)$
and $\ddot{\vec{X}}(t,z)$ in the Euler-Lagrange equation come from
$D_{t}\vec{\Pi}_{t}$, which is given by
\begin{align}
D_{t}\vec{\Pi}_{t}&=-\frac{M_{s}}{\gamma}\int\ud^{2}\vec{r}\;D_{t}\left(\cos\Theta\nabla_{\vec{r}}\Phi\right)\nonumber\\
&=-\frac{M_{s}}{\gamma}\int\ud^{2}\vec{r}\Bigg(\nabla_{\vec{r}}\left[-\dot{X}_{j}
\partial_{j}\Phi+\ddot{X}_{j}\tilde{\partial}_{j}\Phi\right]\cos\Theta\nonumber\\
&\quad+\nabla_{\vec{r}}\Phi\left[-\dot{X}_{j}\partial_{j}\cos\Theta+\ddot{X}_{j}
\tilde{\partial}_{j}\cos\Theta\right]\Bigg)\nonumber\\
&=\hat{e}_{i}M_{ij}\ddot{X}_{j}-\hat{e}_{i}K_{ij}\dot{X}_{j},
\end{align}
where
\begin{align}
\label{tensors}
 M_{ij}&=-\frac{M_{s}}{\gamma}\int\ud^{2}\vec{r}\;\Big[(\partial_{i}
 \tilde{\partial}_{j}\Phi)\cos\Theta+(\partial_{i}\Phi)\tilde{\partial}_{j}\cos\Theta\Big],\nonumber\\
 K_{ij}&=-\frac{M_{s}}
 {\gamma}\int\ud^{2}\vec{r}\;\Big[(\partial_{i}
 \partial_{j}\Phi)\cos\Theta+(\partial_{i}\Phi)\partial_{j}\cos\Theta\Big]
\end{align}
and $\ds \partial_{j}\equiv\nabla_{r_{j}}$, $\ds
\tilde{\partial}_{j}=\nabla_{v_{j}}$.

We want $\hat{e}_{i}K_{ij}\dot{X}_{j}$ to be of the form
$\vec{\rho}_{G}\times\dot{\vec{X}}$ which results in the identity
$\epsilon_{ijk}\rho_{G,j}=K_{ik}$. From this we obtain $\ds
\rho_{G,j}=-\frac{1}{2}\epsilon_{ikj}K_{ik}$, which translates
into the vector form as
\begin{align}
\label{gyro}
\vec{\rho}_{G}&=\frac{M_{s}}{2\gamma}\int\mathrm{d}^{2}\vec{r}\epsilon_{ikj}
\Big[(\partial_{i} \partial_{k}\Phi)\cos\Theta+(
\partial_{i}\Phi)(\partial_{k}\cos\Theta)\Big]
\hat{e}_{j}\nonumber\\
&=\frac{M_{s}}{2\gamma}\int\ud^{2}\vec{r}\;\Big[\big(\nabla_{\vec{r}}\times\nabla_{\vec{r}}\Phi\big)\cos\Theta+\nonumber\\
&\qquad\qquad\qquad\qquad\qquad\qquad\quad\nabla_{\vec{r}}\Phi
\times\nabla_{\vec{r}}\cos\Theta\Big]
\end{align}

To compute the mass tensor and the gyrovector we have to find the
solutions $(\Theta,\Phi)$ of the Landau-Lifshitz equation in the
low dynamics regime that is characterized by the condition
$|\dot{\vec{X}}|\ll1$. In this regime solutions can be expanded as
a perturbative series on the differential speed,
$|\dot{\vec{X}}|$, of the vortex core
\begin{align}
 \Theta(t,z;\vec{r})&=\Theta^{(0)}(z;\vec{r})+\Theta^{(1)}(t,z;\vec{r})+\ldots\nonumber\\
 \Phi(t,z;\vec{r})&=\Phi^{(0)}(z;\vec{r})+\Phi^{(1)}(t,z;\vec{r})+\ldots
\end{align}
Notice that the zero-th order is time independent, otherwise the
gyrovector would depend on time.

The approach that neglects deformation of the vortex core in any
$z$-cross-section of the disk is correct only for weak deviations
of the centerline of the vortex core from the straight line along
the $Z$-axis. We now proceed to the study of the Landau-Lifshitz
equation for the set of variables $(\Theta,\Phi)$ in such weak
bending regime. It can be obtained by applying the variational
principle to the Lagrangian density
\begin{align}
 \tilde{\mathcal{L}}\big[t,z;\Theta,\Phi,\dot{\Theta},\dot{\Phi},&\partial_{z}\Theta,\partial_{z}\Phi\big]=\int\ud^{2}\vec{r}\;\Bigg[\frac{M_{s}}
 {\gamma}(D_{t}\Phi)\cos\Theta\nonumber\\
 &-\mathcal{E}(\Theta,\Phi,\nabla_{\vec{r}}\Theta,
 \nabla_{\vec{r}}\Phi,\partial_{z}\Theta,\partial_{z}\Phi)\Bigg]
\end{align}
Notice that
\begin{align}
\mathcal{E}(\Theta,\Phi,\nabla_{\vec{r}}\Theta,\nabla_{\vec{r}}
\Phi,\partial_{z}\Theta,&\partial_{z}\Phi)=\mathcal{E}_{XY}(\Theta,\Phi,\nabla_{\vec{r}}\Theta,\nabla_{\vec{r}}\Phi)\nonumber\\
&+\mathcal{E}_{el}(\Theta,\Phi,\partial_{z}\Theta,\partial_{z}\Phi)
\end{align}
with
$\mathcal{E}_{XY}(\Theta,\Phi,\nabla_{\vec{r}}\Theta,\nabla_{\vec{r}}\Phi)$
being the sum of the exchange, anisotropy and dipolar energy
responsible for the formation of the vortex, and
\begin{equation}
\mathcal{E}_{el}(\Theta,\Phi,\partial_{z}\Theta,\partial_{z}\Phi)=
A_{eff}\left[(\partial_{z}\Theta)^2+\sin^2\Theta(\partial_{z}\Phi)^2\right]
\end{equation}
being the elastic energy in which $A_{eff}$ is a constant. It
describes contribution of the exchange and dipolar forces to the
elasticity of the vortex line, with the exchange playing a
dominant role. Consequently, with good accuracy, $A_{eff}$ can be
identified with the exchange constant $A$.

The set of dynamical equations for $(\Theta,\Phi)$ is
\begin{eqnarray}
 D_{t}\left(\frac{\delta \tilde{\mathcal{L}}}{\delta(D_{t}\Phi)}\right)+\partial_{z}\left(\frac{\delta\tilde{\mathcal{L}}}
 {\delta(\partial_{z}\Phi)}\right)-
\frac{\delta\tilde{\mathcal{L}}}{\delta\Phi}&=&0\nonumber\\
-2A_{eff}\left[\sin2\Theta\partial_{z}\Theta\partial_{z}\Phi+\sin^2\Theta\partial_{z}^{2}\Phi\right]&+&\nonumber\\
\frac{M_{s}}{\gamma}\frac{d\cos\Theta}{dt}+\frac{\delta\mathcal{E}_{XY}}{\delta\Phi}&=&0
\end{eqnarray}
and
\begin{eqnarray}
 D_{t}\left(\frac{\delta \tilde{\mathcal{L}}}{\delta(D_{t}\Theta)}\right)+\partial_{z}
 \left(\frac{\delta\tilde{\mathcal{L}}}{\delta(\partial_{z}\Theta)}\right)-
\frac{\delta\tilde{\mathcal{L}}}{\delta\Theta}&=&0\nonumber\\
-2A_{eff}\left[\partial_{z}^{2}\Theta-\frac{\sin2\Theta}{2}(\partial_{z}\Phi)^2\right]&+&\nonumber\\
\frac{M_{s}}{\gamma}\sin\Theta\frac{d\Phi}{dt}+\frac{\delta\mathcal{E}_{XY}}{\delta\Theta}&=&0
\end{eqnarray}

Performing a Fourier transform
\begin{eqnarray}
\label{Devel}
 \Phi(t,z;\vec{r})&=&\frac{1}{\sqrt{2\pi}}\int\ud q\;\bar{\Phi}(t,q;\vec{r})e^{iqz}\nonumber\\
 \Theta(t,z;\vec{r})&=&\frac{1}{\sqrt{2\pi}}\int\ud q\;\bar{\Theta}(t,q;\vec{r})e^{iqz}
\end{eqnarray}
we obtain the following set of equations for the pair
$(\bar{\Theta},\bar{\Phi})$:
\begin{align}
\label{Eq1}
 \frac{A_{eff}q^2}{\pi}\Big[\overline{\sin2\Theta}\star\bar{\Theta}\star\bar{\Phi}&+\overline{\sin\Theta}\star\overline{\sin\Theta}\star\bar{\Phi}\Big]+\nonumber\\
 &\frac{M_{s}}{\gamma}\frac{d\;\overline{\cos\Theta}}{dt}+\overline{\frac{\delta\mathcal{E}_{XY}}{\delta\Phi}}=0
 \end{align}
 \begin{align}
\label{Eq2}
 \frac{1}{\sqrt{2\pi}}\frac{M_{s}}{\gamma}&\overline{\sin\Theta}\star\frac{d\overline{\Phi}}{dt}+\overline{\frac{\delta\mathcal{E}_{XY}}{\delta\Theta}}+\nonumber\\
 &2A_{eff}q^2\left[\bar{\Theta}-\frac{\overline{\sin2\Theta}}{4\pi}\star\bar{\Phi}\star\bar{\Phi}\right]=0
\end{align}
where $\star$ means Fourier convolution.

For a small bending of the vortex core, the boundary conditions on
the angle $\Theta$ are the same as in the rigid VC case, i.e.:
$\Theta\simeq0\textrm{ or }\pi$ in the limit
$\tilde{r}\ll\Delta_{0}$ and $\Theta\simeq\pi/2$ in the limit
$\tilde{r}\gg\Delta_{0}$, with $\Delta_{0}=\sqrt{A/M_{s}^2}$ being
the exchange length of the material and where
$\tilde{r}=||\vec{r}-\vec{X}(t,z)||_{2}$ is the radial distance
from the VC center at any height $z$. Considering this two limits
Eq. \eqref{Eq1} becomes:
\begin{itemize}
 \item[$\bullet$] Limit $\tilde{r}\ll\Delta_{0}$.
     In this case, $\sin\Theta\simeq0$ and thus $\sin2\Theta\partial_{z}\Theta\partial_{z}\Phi+\sin^2\Theta\partial_{z}^{2}\Phi\simeq0$. So we have the following equation in the Fourier space
      \begin{equation}
       \frac{M_{s}}{\gamma}\frac{d\;\overline{\cos\Theta}}{dt}+\overline{\frac{\delta\mathcal{E}_{XY}}{\delta\Phi}}=0
      \end{equation}
 \item[$\bullet$] Limit $\tilde{r}\gg\Delta_{0}$.
     In this case, $\sin\Theta\simeq1$ and thus $\sin2\Theta\partial_{z}\Theta\partial_{z}\Phi+\sin^2\Theta\partial_{z}^{2}\Phi\simeq\partial_{z}^2\Phi$.
     So we have the equation
     \begin{equation}
      \frac{M_{s}}{\gamma}\frac{d\;\overline{\cos\Theta}}{dt}+2A_{eff}q^2\bar{\Phi}+\overline{\frac{\delta\mathcal{E}_{XY}}{\delta\Phi}}=0
     \end{equation}
\end{itemize}
Notice that in both limits $\sin2\Theta\simeq0$ and so
$\partial_{z}^{2}\Theta-\frac{\sin2\Theta}{2}(\partial_{z}\Phi)^2\simeq\partial_{z}^{2}\Theta$.
Consequently, in the Fourier space Eq. \eqref{Eq2} becomes
\begin{equation}
 \frac{1}{\sqrt{2\pi}}\frac{M_{s}}{\gamma}\overline{\sin\Theta}\star\frac{d\overline{\Phi}}{dt}+2A_{eff}q^2\bar{\Theta}+\overline{\frac{\delta\mathcal{E}_{XY}}{\delta\Theta}}=0
 \end{equation}

Finally, in the limit of weak bending ($A_{eff}q^2\ll1$), we can
neglect the terms of the form $2A_{eff}q^2\bar{\xi}$ in the above
equations. In doing so, we recover the standard Landau-Lifshitz
equations for $(\Theta,\Phi)$ at any $z$ layer, with the VC center
depending on the value of $z$. Introducing now the perturbative
series \eqref{Devel} into the Landau-Lifshitz equation and
splitting it into $O(|\dot{\vec{X}}|^{n})$ terms, we obtain the
equations of motion for the $\Phi^{(n)}/\Theta^{(n)}$ terms. In
the case of the zero-th and first order terms, we recover the
static solution and the first perturbative solution for the rigid
vortex (see Section VI). For the particular case of the zero-th
order we obtain
\begin{align}
\label{zeroth}
\Phi_{0}(x,y)&=n_v\tan^{-1}(y-y_{v}/x-x_{v})\nonumber\\
\cos\Theta_{0}(\tilde{r})&= \left\{\begin{array}{lcc}
            p\left(1-C_{1}\left(\frac{\tilde{r}}{\Delta_{0}}\right)^2\right)
            & & \tilde{r} \ll\Delta_{0}\\
        C_{2}\left(\frac{\Delta_{0}}{\tilde{r}}\right)^{1/2}\exp(-\tilde{r}/\Delta_{0})
        & & \tilde{r}\gg\Delta_{0}
           \end{array}\right.
\end{align}
where $n_v = \pm 1$ is the vorticity of the magnetization of the
disk and $C_{1},C_{2}$ are constants that can be obtained by
imposing the smoothness condition on $\cos\Theta_{0}$ at
$\tilde{r}=\Delta_{0}$ up to its first derivative. The
corresponding values are $C_{1}=\frac{3}{7}$ and
$C_{2}=\frac{4}{7}pe$. From all this we straightforwardly deduce
that
\begin{eqnarray}
\label{Phi_prop} \nabla_{\vec{r}}\times\nabla_{\vec{r}}\Phi_{0}& =
& 2\pi n_v \delta^{(2)}
\left(\vec{r}-\vec{X}(t,z)\right)\hat{e}_{z}\nonumber\\
\nabla_{\vec{r}}^{2}\Phi_{0} &=& 0
\end{eqnarray}

\section{Elastic Thiele equation}

We now proceed to the computation of the gyrovector and the mass
density tensor. Using \eqref{Phi_prop} we obtain that the first
term of Eq. \eqref{gyro} equals $(\pi n_v p
M_{s}/\gamma)\hat{e}_{z}$, where $p=\cos\Theta(\vec{0})=\pm1$
defines the direction of the polarization of the vortex core
($\Theta(\vec{0})=0$ or $\pi$). The second term is evaluated at
the zero-th order of the perturbative expansion of the angular
coordinates in the low dynamics regime, Eq. \eqref{zeroth}, taking
into account that in the weak bending regime the deformation of
the vortex core is small and one can consider $\tilde{r}\simeq r$
because $||\vec{X}(t,z)||_{2}\ll1$. In doing so we obtain $(\pi
n_v p M_{s}/\gamma)\hat{e}_{z}$ again\cite{Huber}. Thus the
gyrovector becomes $\vec{\rho}_{G}=\rho_{G}pn_v\hat{e}_{z}$ with
\begin{equation}\label{rho-G}
\rho_{G}=2\pi M_{s}/\gamma
\end{equation}
Notice that $\vec{\rho}_{G}$ is the gyrovector linear density as
compared to the gyrovector in the Thiele equation for a rigid
vortex\cite{Thiele}.

Computation of the mass density tensor will be performed in
Section VI. For circular polarized disks we show that this tensor
reduces to a scalar, $M_{ij}=\rho_{M}\delta_{ij}$, with the vortex
core mass density given by
$\ds\rho_{M}=\frac{1}{4\gamma^2}\ln(R/\Delta_{0})$, where $R$ is
the radius of the disk. Only $\omega(\vec{X},\partial_{z}\vec{X})$
contributes to the partial derivative
$\delta\tilde{\mathcal{L}}/\delta\vec{X}$ in the slow dynamics
regime, because $\vec{\Pi}_{t}=\rho_{M}\dot{\vec{X}}$ and so the
term $\dot{\vec{X}}\cdot\vec{\Pi}_{t}$ equals
$\rho_{M}\dot{\vec{X}}^2$. Consequently, the generalized Thiele
equation becomes
\begin{equation}
\label{Thiele}
\rho_{M}\ddot{\vec{X}}(t,z)+\dot{\vec{X}}(t,z)\times\vec{\rho}_{G}
+\partial_{z}\vec{\Pi}_{z}+\nabla_{\vec{X}}\omega=0
\end{equation}

The linear energy density $\omega(\vec{X},\partial_{z}\vec{X})$ is
the sum of the magnetostatic and exchange contributions in the
$z$-cross-section, $\omega_{XY}(\vec{X})$, and an elastic
contribution due to the deformation of the vortex core line,
$\omega_{el}(\partial_{z}\vec{X})$. Zeeman contribution will be
considered later. The dependence on the vortex core coordinates on
the $\omega_{XY}(\vec{X})$ term for small displacements
is\cite{Guslienko1, Guslienko2}
\begin{equation}
\omega_{XY}(\vec{X})=\frac{1}{2}\rho_{M}\omega_{M}^2\epsilon_{0}\vec{X}^2,
\end{equation}
where $\ds \omega_{M}=\rho_{G}/\rho_{M}$ is the characteristic
frequency of the system and $\epsilon_{0}=\omega_{G}/\omega_{M}$
is a dimensionless parameter. Recall that the conventional
gyrofrequency $\omega_{G}$ is defined as\cite{Guslienko2,
Guslienko3}
\begin{equation}
\label{gyrofreq}
\omega_{G}=\frac{\omega_{XY}^{''}(\vec{X}=\vec{0})}{\rho_{G}}\simeq\frac{20}{9}\gamma
M_{s}\beta
\end{equation}
where $\beta=L/R$ is the ratio of the thickness and the radius of the disk. This last expression is valid in the limit $\beta\ll 1$.

From the continuous spin-field model we know that
\begin{equation}
\ds
\omega_{el}(\partial_{z}\vec{X})=A_{eff}\int\ud^{2}\vec{r}\left[(\partial_{z}\Theta)^2
+\sin^2\Theta(\partial_{z}\Phi)^2\right]
\end{equation}
Noticing that
$\partial_{z}\Theta=-\nabla_{\vec{r}}\Theta\cdot\partial_{z}\vec{X}$
and
$\partial_{z}\Phi=-\nabla_{\vec{r}}\Phi\cdot\partial_{z}\vec{X}$,
and taking into account the vector identity,
$(\vec{A}\times\vec{B})\cdot(\vec{C}\times\vec{D})=(\vec{A}\cdot\vec{C})
(\vec{B}\cdot\vec{D})-(\vec{A}\cdot\vec{D})(\vec{B}\cdot\vec{C})$,
with either
$\vec{A}=\vec{D}=\nabla_{\vec{r}}\Theta,\;\vec{B}=\vec{C}=-\partial_{z}\vec{X}$
or
$\vec{A}=\vec{D}=\nabla_{\vec{r}}\Phi,\;\vec{B}=\vec{C}=-\partial_{z}\vec{X}$,
we obtain the following relations
\begin{align}
\left(\nabla_{\vec{r}}\Theta\cdot\partial_{z}\vec{X}\right)^{2}&=\left(\nabla_{\vec{r}}\Theta\right)^{2}\left(\partial_{z}\vec{X}\right)^{2}-\left(\nabla_{\vec{r}}\Theta\times\partial_{z}\vec{X}\right)^{2}\\
\left(\nabla_{\vec{r}}\Phi\cdot\partial_{z}\vec{X}\right)^{2}&=\left(\nabla_{\vec{r}}\Phi\right)^{2}\left(\partial_{z}\vec{X}\right)^{2}-\left(\nabla_{\vec{r}}\Phi\times\partial_{z}\vec{X}\right)^{2}
\end{align}
The main contribution to the integral comes from the zero-th order
in the perturbative expansion \eqref{Devel}. Notice that
$\Theta_{0}(\vec{r})=\Theta_{0}(\tilde{r})$ and that
$\nabla_{\vec{r}}\Phi_{0}=\frac{n_{v}}{\tilde{r}}\hat{e}_{\phi}$.
As discussed before, in the weak bending regime we can use
approximation $\tilde{r}\simeq r$, so that
$\nabla_{\vec{r}}\Phi_{0}\times\partial_{z}\vec{X}=
-\frac{n_{v}}{r}\partial_{z}X_{r}\hat{e}_{z}$ and
$\nabla_{\vec{r}}\Theta_{0}\times\partial_{z}\vec{X}=
\frac{d\Theta_{0}}{dr}\partial_{z}X_{\phi}\hat{e}_{z}$, where
$X_{r}=\hat{e}_{r}\cdot\vec{X}=x\cos\theta+y\sin\theta$ and
$X_{\phi}=\hat{e}_{\phi}\cdot\vec{X}=-x\sin\theta+y\cos\theta$.
The elastic energy density finally becomes
\begin{align}
\omega_{el}(\partial_{z}\vec{X})=&A_{eff}\int\ud^{2}\vec{r}
\left[(\nabla_{\vec{r}}\Theta_{0})^{2}+\frac{\sin^{2}
\Theta_{0}}{r^{2}}\right]\left(\frac{\partial\vec{X}}
{\partial z}\right)^{2}\nonumber\\
&-A_{eff}\int\ud^{2}\vec{r}\left(\frac{d\Theta_{0}}{dr}\right)^{2}
\left(\frac{\partial X_{\phi}}{\partial z}\right)^{2}\nonumber\\
&-A_{eff}\int\ud^{2}\vec{r}\;\frac{\sin^{2}\Theta_{0}}{r^2}\left(\frac{\partial X_{r}}
{\partial z}\right)^{2}\nonumber\\
=&\pi A_{eff}\int r\ud r\left[\left(\frac{d\Theta_{0}}{dr}
\right)^{2}+\frac{\sin^{2}\Theta_{0}}{r^{2}}\right]
\left(\frac{\partial\vec{X}}{\partial z}\right)^{2}
\end{align}
where the angular dependence of $(\partial_{z}X_{r})^{2}$ and
$(\partial_{z}X_{\phi})^{2}$ has been integrated on $\theta$. We
can recast this energy density as
$\ds\omega_{el}(\partial_{z}\vec{X})=\frac{1}{2}\lambda
\left(\frac{\partial\vec{X}}{\partial z}\right)^2$, where
$\lambda$ is the elastic constant given by
\begin{equation}
\label{lambda}
\lambda=2\pi A_{eff}\int r\ud r\left[\left(\frac{d\Theta_{0}}{dr}\right)^{2}+\frac{\sin^{2}\Theta_{0}}{r^{2}}\right]
\end{equation}
Making use of the variable $m_{0}(r)=\cos\Theta_{0}(r)$, we can rewrite the above equation as
\begin{equation}
\lambda=2\pi A_{eff} \int r\ud r\left[\frac{1}{1-m_{0}^{2}}\left(\frac{dm_{0}}{dr}\right)^{2}+\frac{1-m_{0}^{2}}{r^{2}}\right]
\end{equation}
and using the spatial dependence \eqref{zeroth}, we get
\begin{align}
\label{integrand}
&\frac{1}{1-m_{0}^{2}}\left(\frac{dm_{0}}{dr}\right)^{2}+\frac{1-m_{0}^{2}}{r^{2}}
=\nonumber\\
&\left\{\begin{array}{lcc}
\frac{\Delta_{0}^{2}}{2C_{1} r^{2}}\left(\frac{2pC_{1}}{\Delta_{0}^{2}}r\right)^{2}
+\frac{2C_{1}r^{2}}{\Delta_{0}^{2}}\frac{1}{r^{2}}=\frac{4C_{1}}{\Delta_{0}^{2}}
& & r\ll\Delta_{0}\\
\frac{1}{\Delta_{0}^{2}}m_{0}^{2}+\frac{1}{r^{2}}=\frac{C_{2}^{2}}{\Delta_{0}r}
\exp{(-2r/\Delta_{0})}+\frac{1}{r^{2}}& & r\gg\Delta_{0}
\end{array}\right.
\end{align}
Recalling that $C_{1}=\frac{3}{7}$ and $C_{2}=\frac{4}{7}pe$ and
computing the integral \eqref{lambda} by splitting into two
regions, $[0,\Delta_{0}]$ and $[\Delta_{0},R]$, we obtain
\begin{equation}
\lambda=2\pi A_{eff}\left(\frac{50}{49}+\ln(R/\Delta_{0})\right)
\end{equation}
In the limit $R\gg \Delta_{0}$ the logarithmic term dominates and
$\lambda$ becomes
\begin{equation}\label{lambda-A}
\lambda=2\pi A_{eff}\ln(R/\Delta_{0})
\end{equation}

Finally, for the total energy density we obtain
\begin{equation}
\label{Potential}
\omega(\vec{X},\partial_{z}\vec{X})=\frac{1}{2}\rho_{M}\omega_{M}^2\epsilon_{0}
\vec{X}^2+\frac{1}{2}\lambda\left(\frac{\partial\vec{X}}{\partial
z}\right)^2
\end{equation}
and thus the generalized Thiele equation for an elastic vortex
core line becomes
\begin{equation}
\label{Thiele2}
\rho_{M}\ddot{\vec{X}}(t,z)-\lambda\partial^{2}_{z}\vec{X}(t,z)
+\dot{\vec{X}}(t,z)\times\vec{\rho}_{G}+\rho_{M}\omega_{M}^2\epsilon_{0}\vec{X}(t,z)=0
\end{equation}

\section{Spin waves in the vortex core}
Introducing the complex variable $\chi=x-iy$ we can recast Eq.
\eqref{Thiele2} as the following complex partial differential
equation:
\begin{equation}
 \label{CThiele}
\rho_{M}\ddot{\chi}-\lambda\partial^{2}_{z}\chi+i\rho_{G}\dot{\chi}+\rho_{M}
\omega_{M}^2\epsilon_{0}\chi=0
\end{equation}
Let $\chi_{0}(z)$ be the equilibrium complex center of the
straight vortex core line. In the presence of the wave it gets
perturbed and becomes $\chi(t,z)=\chi_{0}(z)+\chi_{w}(t,z)$, with
$||\chi_{w}||_{z}\ll||\chi_{0}||_{z}$. Switching to the Fourier
transform,
\begin{equation}
\chi_{w}(t,z)=\frac{1}{2\pi}\int\ud\omega\ud
q\chi_{w}(\omega,q)e^{i(\omega t-qz)},
\end{equation}
we obtain the following equation for $\chi_{w}(\omega,q)$:
\begin{equation}
 \label{FCThiele}
\left[-\rho_{M}\omega^2+\lambda
q^2-\rho_{G}\omega+\rho_{M}\omega_{M}^2\epsilon_{0}\right]\chi_{w}(\omega,q)=0
\end{equation}
For non-zero amplitude of the wave the expression in the square
parenthesis must vanish. This determines the spectrum of the
waves:
\begin{equation}
 \label{spectrum}
 \rho_{M}\omega^2-\lambda q^2+\rho_{G}\omega-\rho_{M}\omega_{M}^2\epsilon_{0}=0
\end{equation}

At $\rho_{M}\neq0$ one can normalize Eq. \eqref{spectrum} to get
 \begin{equation}
  \omega^2+\omega_{M}\omega-\omega_{M}^2\left(\epsilon_{0}+
  \frac{\lambda}{\rho_{M}\omega_{M}^{2}}q^2\right)=0
 \end{equation}
 Solving this equation we obtain the spectrum of vortex core
 excitations:
\begin{equation}\label{waves}
 \omega_{\pm}(q)=\frac{\omega_{M}}{2}\left[\sqrt{(1+4\epsilon_{0})+
 \frac{4\lambda q^2}{\rho_{M}\omega_{M}^2}}\pm1\right]
\end{equation}
In the weak bending regime we have
$4\lambda q^2/\rho_{M}\omega_{M}^2\ll1$ and so we can expand the
square root and obtain the following expression for the
frequencies
\begin{equation}
 \omega_{\pm}(q)=\frac{\omega_{M}}{2}\left(\sqrt{1+4\epsilon_{0}}\pm1\right)
 +\frac{1}{\sqrt{1+4\epsilon_{0}}}\frac{\lambda}{\rho_G}q^2
\end{equation}
where we have used the relation $\rho_M\omega_M = \rho_G$.

As will be shown in Section IV the parameter $\epsilon_0 =
\omega_G/\omega_M$ is normally small due to the smallness of
$\beta = L/R$. Consequently
\begin{eqnarray}
\omega_-(q) & \approx &  \omega_{G}+\frac{\lambda}{\rho_{G}}q^2
\\
\omega_+(q) & \approx & \omega_M + \frac{\lambda}{\rho_{G}}q^2
\end{eqnarray}
With the help of Eqs.\ (\ref{rho-G}) and (\ref{lambda-A}) with
$A_{eff} \approx A = M_s^2\Delta_0^2$ the above equations can be
written in a transparent form:
\begin{eqnarray}
\omega_-(q) &= & \omega_{G}+\gamma M_s
(q\Delta_0)^2\ln(R/\Delta_0) \\
\omega_+(q) &= & \omega_{M}+\gamma M_s
(q\Delta_0)^2\ln(R/\Delta_0)
\end{eqnarray}
Note that the weak bending regime corresponds to $q\Delta_0 \ll
1$.

\section{Quantum mechanics of the excitations in the vortex core}
In this section we will show that excitations of the vortex core
can be also obtained in a rather non-trivial way from the quantum
theory as well. This problem is interesting on its own as it turns
out to be equivalent to the problem of quantum excitations of a
charged string confined in a parabolic potential and subjected to
the magnetic field.

It is straightforward to prove that the generalized Thiele
equation \eqref{Thiele2} is the Euler-Lagrange equation associated
with the following effective Lagrangian density that can be
derived from Eq.\ (\ref{Lagdensity})
\begin{equation}
 \tilde{\mathcal{L}}(t,z;\vec{X},\dot{\vec{X}},\partial_{z}\vec{X})
 =\frac{1}{2}\rho_{M}\dot{\vec{X}}^2+\dot{\vec{X}}
 \cdot\vec{A}_{\rho_{G}}-\omega(\vec{X},\partial_{z}\vec{X})
\end{equation}
where $\vec{A}_{\rho_{G}}$ is the gyrovector potential satisfying
$\nabla_{\vec{X}}\times\vec{A}_{\rho_{G}}=-\vec{\rho}_{G}$. Thus
the total Lagrangian becomes
\begin{equation}
\label{Lagrangian3}
 \mathcal{L}=\int\ud z\;\tilde{\mathcal{L}}=\int \ud z\;\left[\frac{1}{2}\rho_{M}
\dot{\vec{X}}^2+\dot{\vec{X}}\cdot\vec{A}_{\rho_{G}}-\omega(\vec{X},\partial_{z}\vec{X})\right]
\end{equation}
Noticing that
$\ds\left\{\varphi_{n}(z)\right\}_{n\in\mathbf{N}}=
\left\{\sqrt{\frac{2}{L}}\sin\left(q_{n}z\right)\right\}_{n\in\mathbf{N}}$,
with $\ds q_{n}=\frac{2\pi}{L}n$, is a Hilbert basis of the
function subspace
$\mathcal{W}=\{\varphi\in\mathcal{L}^{2}(0,L),\;\varphi(0)
=\varphi(L)=0\}$, we
can expand $\vec{X}$ as
\begin{equation}
 \vec{X}(t,z)=\vec{X}_{0}(t)+\sum_{n}\vec{X}_{n}(t)\varphi_{n}(z),
\end{equation}
where $\vec{X}_{0}(t)$ is the center of the undisturbed vortex and
$\vec{X}_{n}(t)=<\vec{X}(t,z),\varphi_{n}(z)>_{\mathcal{L}^2(0,L)}$.
Introducing this expansion in Eq. \eqref{Lagrangian3} and taking
into account the orthonormality of the Hilbert basis (and its
spatial derivatives) we obtain the following identity:
\begin{align}
&\mathcal{L}\Bigg(t,\left\{\vec{X}_{n}\right\}_{n\in\mathbf{Z}^{+}},\left\{\dot{\vec{X}}_{n}\right\}_{n\in\mathbf{Z}^{+}}\Bigg)=\nonumber\\
&\left[\frac{1}{2}M\dot{\vec{X}}_{0}^{2}+\dot{\vec{X}}_{0}\cdot\vec{A}_{0}-\frac{1}{2}M\omega_{M}^{2}\epsilon_{0}\vec{X}_{0}^{2}\right]+\nonumber\\
&\sum_{n>0}\left[\frac{1}{2}\rho_{M}\dot{\vec{X}}_{n}^{2}+\dot{\vec{X}}_{n}\cdot\vec{A}_{n}-\frac{1}{2}\rho_{M}\omega_{M}^{2}\epsilon_{0}\vec{X}_{n}^{2}-\frac{1}{2}\lambda
q_{n}^{2}\vec{X}_{n}^{2}\right]
\end{align}
where $\mathbf{Z}^{+}=\{0\}\cup\mathbf{N}$, $M=\rho_{M}L$ is the
total mass of the rigid vortex line and $\vec{A}_{n}$ is the
gyrovector potential associated to the $n$-th coordinate
$\vec{X}_{n}$, which satisfies
$\nabla_{\vec{X}_{0}}\times\vec{A}_{0}=-\vec{G}$ and
$\nabla_{\vec{X}_{n}}\times\vec{A}_{n}=-\vec{\rho}_{G},\; n>0$,
with $\vec{G}=\vec{\rho}_{G}L$ being the gyrovector of the rigid
vortex.

Applying the Laguerre transformation to the above Lagrangian we
obtain the following expression for the Hamiltonian
\begin{align}
\label{Hamiltonian}
&\mathcal{H}\left(t,\left\{\vec{X}_{n}\right\}_{n\in\mathbf{Z}^{+}},\left\{\vec{\Pi}_{n}\right\}_{n\in\mathbf{Z}^{+}}\right)=\nonumber\\
&\left[\frac{1}{2M}(\vec{\Pi}_{0}-\vec{A}_{0})^{2}+\frac{1}{2}M\omega_{M}^{2}\epsilon_{0}\vec{X}_{0}^{2}\right]+\nonumber\\
&\sum_{n>0}\left[\frac{1}{2\rho_{M}}(\vec{\Pi}_{n}-\vec{A}_{n})^{2}+\frac{1}{2}\rho_{M}\omega_{M}^{2}\left(\epsilon_{0}+\frac{\lambda}{\rho_{M}\omega_{M}^{2}}
q_{n}^{2}\right)\vec{X}_{n}^{2}\right]
\end{align}
where $\vec{\Pi}_{0}=M\dot{\vec{X}}_{0}+\vec{A}_{0}$ and
$\vec{\Pi}_{n}=\rho_{M}\dot{\vec{X}}_{n}+\vec{A}_{n},\;n>0$ are
the corresponding canonical momenta. Notice that Eq.
\eqref{Hamiltonian} shows that $\mathcal{H}$ splits into the
direct sum $\oplus_{m\in\mathbf{Z}^{+}}\mathcal{H}_{m}$, with
$\mathcal{H}_{m}$ being the Hamiltonian defined over the phase
space $(\vec{X}_{m},\vec{\Pi}_{m})$. It has a structure of the
form
\begin{equation}
 \label{Hamiltonian2}
  \mathcal{H'}=\frac{1}{2\eta}(\vec{\Pi}-\vec{A})^{2}+\frac{1}{2}\eta\omega_{M}^{2}\xi\vec{X}^{2},
\end{equation}
where $(\vec{X},\vec{\Pi})$ are the canonically conjugate
variables, $\eta,\xi$ are constants and $\vec{A}$ is the
gyrovector satisfying
$\nabla_{\vec{X}}\times\vec{A}=-\chi\hat{z}$, with $\chi$ being a
constant. It is important to point out that $\chi/\eta=\omega_{M}$
in all cases.

From now on we consider the case of the vortex core of a non-zero
mass, ($\eta\neq0$). It is convenient to choose a "symmetric
gauge" given by
\begin{equation}
\label{Gauge}
\vec{A}=\frac{1}{2}(-\chi\hat{z})\times\vec{X}=\frac{\chi
y}{2}\hat{x}-\frac{\chi x}{2}\hat{y}
\end{equation}
Firstly, we define the kinetic momentum operators as
$\vec{p}=\eta\dot{\vec{X}}$, so that $\vec{\Pi}=\vec{p}+\vec{A}$.
Notice the following non vanishing commutators
\begin{equation}
\label{pxpy} \left [p_{j},p_{k}\right]=-i\hbar \chi \epsilon_{jk}
\qquad j,k\in\{x,y\},
\end{equation}
where $\epsilon_{jk}$ is the antisymmetric tensor $\epsilon_{xy} =
-\epsilon_{yx} = 1$. Secondly, we introduce the operators
\begin{align}
\label{Aoperators}
a&=\sqrt{\frac{1}{2\hbar \chi}}\left(p_{y}+ip_{x}\right)\nonumber\\
a^{\dagger}&=\sqrt{\frac{1}{2\hbar
\chi}}\left(p_{y}-ip_{x}\right)
\end{align}
which satisfy standard commutation relations for Bose operators,
$\ds \left[a,a^{\dagger}\right]=1$. The number operator
$N_{a}=a^{\dagger}a$ satisfies commutation relations
$\left[N_{a},a\right]=-a,\quad\left[N_{a},a^{\dagger}\right]=a^{\dagger}$
and we have the identity
\begin{equation}
\frac{1}{2\eta}\left(\vec{\Pi}-\vec{A}\right)^2=\hbar\omega_{M}\left(N_{a}+\frac{1}{2}\right),
\end{equation}

In analogy with the case of a charged particle in the
electromagnetic field\cite{Zak}, we obtain that the gyrotropic
translational group is generated by $\vec{T}=\vec{\Pi}+\vec{A}$,
\begin{equation}
\label{Trans gen} T_{x}=p_{x}+\chi y,\qquad\qquad T_{y}=p_{y}-\chi
x
\end{equation}
which satisfies the following commutation relations,
\begin{equation}
\label{Comm trans gen}
\left[T_{j},p_{k}\right]=0,\qquad\left[T_{j},T_{k}\right]=i\hbar
\chi\epsilon_{jk},\quad j,k\in\{x,y\}
\end{equation}
Now we introduce another set of Bose operators
\begin{align}
\label{Boperators}
b&=\sqrt{\frac{1}{2\hbar \chi}}\left(T_{y}-iT_{x}\right)\nonumber\\
b^{\dagger}&=\sqrt{\frac{1}{2\hbar
\chi}}\left(T_{y}+iT_{x}\right),
\end{align}
which satisfy commutation relations
\begin{equation*}
\left[b,b^{\dagger}\right]=1,\quad\left[M_{b},b\right]=-b,\quad\left[M_{b},b^{\dagger}\right]=b^{\dagger},
\end{equation*}
where $M_{b}=b^{\dagger}b$ is the corresponding number operator.
Notice that the commutation relations
$\left[a,b\right]=\left[a,b^{\dagger}\right]=0$ also hold.

Coordinates $x$ and $y$ can be expressed in terms of the above
Bose operators:
\begin{align}
\label{xy}
x&=\frac{1}{\chi}\left(p_{y}-T_{y}\right)\\
y&=-\frac{1}{\chi}\left(p_{x}-T_{x}\right)\nonumber,
\end{align}
so that
\begin{equation}
\label{r2}
\frac{1}{2}\eta\omega_{M}^2\left(x^2+y^2\right)=\hbar\omega_{M}\left(N_{a}+M_{b}-ab-a^{\dagger}b^{\dagger}+1\right).
\end{equation}
Consequently, the Hamiltonian \eqref{Hamiltonian2} becomes
\begin{equation}
\label{Hamiltonian3}
\mathcal{H'}=\hbar\omega_{M}\left[(1+\xi)N_{a}+\xi
M_{b}-\xi(ab+a^{\dagger}b^{\dagger})+\xi+\frac{1}{2}\right]
\end{equation}
It can be diagonalized with the help of Bogoliubov transformations
\begin{equation}
\label{Bogoliubov} \bar{\alpha}=ua-vb^{\dagger},\qquad
\bar{\beta}=ub-va^{\dagger}
\end{equation}
with $u,v$ being real numbers. These new operators satisfy Bose
commutation relations if $u^2-v^2=1$. Substituting the above
equations into Eq. \eqref{Hamiltonian3} we obtain
\begin{align}
\label{Hamiltonian4}
\mathcal{H'}=\hbar\omega_{M}&\Bigg[\bar{\alpha}^{\dagger}\bar{\alpha}(u^2(1+\xi)+\xi v^2-2\xi uv)+\\
&\bar{\beta}^{\dagger}\bar{\beta}(v^2(1+\xi)+\xi u^2-2\xi uv)+\nonumber\\
&(\bar{\alpha}^{\dagger}\bar{\beta}^{\dagger}+\bar{\alpha}\bar{\beta})\left(uv(1+2\xi)-\xi(u^2 + v^2)\right)\nonumber\\
&+\left(v^2(1+2\xi)-2\xi uv+(\xi+1/2)\right)\Bigg]\nonumber
\end{align}
To get a Hamiltonian in the oscillator form, the coefficient
related to ($\bar{\alpha}^{\dagger}\bar{\beta}^{\dagger}+
\bar{\alpha}\bar{\beta}$) should be zero, which requires
\begin{equation}
uv(1+2\xi)-\xi(u^2+v^2)=0
\end{equation}
The solution is $u=\cosh(\theta),\; v=\sinh(\theta)$,
\begin{equation}
\tanh(2\theta)=\frac{2\xi}{1+2\xi}
\end{equation}
Finally, the coefficients of the terms
$\bar{\alpha}^{\dagger}\bar{\alpha}$ and
$\bar{\beta}^{\dagger}\bar{\beta}$ become
\begin{align}
u^2(1+\xi)+\xi v^2-2\xi uv&=\frac{1}{2}\left[\frac{1+2\xi}{\cosh(2\theta)}+1\right]\nonumber\\
&=\frac{1}{2}\left[\sqrt{1+4\xi}+1\right]\nonumber\\
v^2(1+\xi)+\xi u^2-2\xi
uv&=\frac{1}{2}\left[\frac{1+2\xi}{\cosh(2\theta)}-1\right]\nonumber\\
&=\frac{1}{2}\left[\sqrt{1+4\xi}-1\right]
\end{align}
and, consequently, the Hamiltonian in the second quantization
formalism becomes
\begin{equation}
\label{Hamiltonian5}
\mathcal{H'}=\hbar\omega_{+}\left(\bar{\alpha}^{\dagger}\bar{\alpha}+\frac{1}{2}\right)+\hbar\omega_{-}\left(\bar{\beta}^{\dagger}\bar{\beta}+\frac{1}{2}\right)
\end{equation}
where
\begin{align}
\label{freq}
\omega_{\pm}&=\frac{1}{2}\left[\sqrt{1+4\xi}\pm1\right]\omega_{M}
\end{align}

Noticing that for any $n\in\mathbf{Z}^{+}$ we have
$\xi=\epsilon_{0}+\frac{\lambda}{\rho_{M}\omega_{M}^{2}} q_{n}^{2}$,
the second quantization procedure yields the following form of
Hamiltonian \eqref{Hamiltonian}
\begin{equation}
\mathcal{H}=\sum_{n\geq0}\hbar\omega^{+}_{n}\left(\bar{\alpha}_{n}^{\dagger}\bar{\alpha}_{n}+\frac{1}{2}\right)+\sum_{n\geq0}\hbar\omega^{-}_{n}\left(\bar{\beta}_{n}^{\dagger}\bar{\beta}_{n}+\frac{1}{2}\right)
\end{equation}
where $\omega^{\pm}_{n}$ are the eigenfrequencies of the vortex
state given by
\begin{equation}
\label{freq2}
\omega_{n}^{\pm}=\frac{1}{2}\left[\sqrt{(1+4\epsilon_{0})+\frac{4\lambda}{\rho_{M}\omega_{M}^{2}}
q_{n}^{2}}\pm1\right]\omega_{M}
\end{equation}
which coincides with (\ref{waves}).

\section{Computation of the vortex mass}
As it has been discussed in Section II, to calculate the vortex mass density tensor (see Eq. \eqref{tensors}) we
need to find solution $(\Phi(\vec{r},t),\Theta(\vec{r},t))$ of the
Landau-Lifshitz equation in the slow dynamics regime, i.e. in the
first order on $|\dot{\vec{X}}|$. A more convenient set of
variables for this problem is the pair $(\Phi,m)$, where $\ds
m\equiv m_{z}=\frac{M_{z}}{M_{s}}=\cos\Theta$ is the projection of
the magnetic moment onto the $z$ axis. Notice that Landau-Lifshitz
equation can be recast as the set of equations
\begin{align}
\label{LL2}
 \frac{d\Phi}{dt}&=\frac{\gamma}{M_{s}}\frac{\delta\mathcal{E}}{\delta m}\nonumber\\
 \frac{dm}{dt}&=-\frac{\gamma}{M_{s}}\frac{\delta\mathcal{E}}{\delta\Phi}
\end{align}
The total energy $\mathcal{E}(\Phi,m)$ splits into the sum
\begin{align}
 \mathcal{E}(\Phi,m)&=\mathcal{E}_{ex}(\Phi,m)+\mathcal{E}_{an}(\Phi,m)
 +\mathcal{E}_{demag}(\Phi,m)\nonumber\\
&=A\left[(\nabla\Theta)^2+\sin^2\Theta(\nabla\Phi)^2\right]\nonumber\\
&\quad-K_{\shortparallel}\frac{M_{x}^2}{M_{s}^2}+K_{\perp}\frac{M_{z}^2}{M_{s}^2}
-\frac{1}{2}\vec{M}\cdot\vec{H}_{d}\nonumber\\
&=A\left[\frac{1}{1-m^2}(\nabla m)^2+(1-m^2)(\nabla\Phi)^2\right]\nonumber\\
&-K_{\shortparallel}\cos^2\Phi(1-m^2)+K_{\perp}m^2-\frac{1}{2}
\vec{M}\cdot\vec{H}_{d}
\end{align}
with $A$ being the exchange constant,
$K_{\shortparallel},\;K_{\perp}$ being the anisotropy constants
and $\vec{H}_{d}$ being the demagnetizing field. Recall that
$\vec{H}_{d}(\vec{r})=-\nabla\Phi_{d}(\vec{r})$, with
$\nabla^{2}\Phi_{d}(\vec{r})=-4\pi\rho_{d}$ and
$\rho_{d}=-\nabla\cdot\vec{M}$. Equivalently\cite{CT-lectures},
\begin{align}
 \Phi_{d}(\vec{r})&=\int_{V}\ud^{3}\vec{r}' \vec{M}(\vec{r}')\cdot\nabla'
 \left(\frac{1}{|\vec{r}-\vec{r}'|}\right)\nonumber\\
&=-\int_{V}\ud^{3}\vec{r}'\frac{\nabla'\cdot\vec{M}(\vec{r}')}{|\vec{r}-\vec{r}'|}
+\int_{\partial V}\ud\vec{S}'\cdot\frac{\vec{M}(\vec{r}')}{|\vec{r}-\vec{r}'|}\nonumber\\
&=\int_{V}\ud^{3}\vec{r}'\frac{\rho_{d}(\vec{r}')}{|\vec{r}-\vec{r}'|}
+\int_{\partial V}\ud^{2}\vec{r}'\frac{\sigma_{d}(\vec{r}')}{|\vec{r}-\vec{r}'|}
\end{align}
with $\sigma_{d}=\vec{M}\cdot\vec{n}$ being the effective surface
``charge'' density and $V$ being the volume of the system.
Consequently, the demagnetizing energy can be written as
\begin{equation}
\mathcal{E}_{demag}=\frac{1}{2}\int_{V}\ud^{3}\vec{r}\rho_{d}(\vec{r})
\Phi_{d}(\vec{r})+\frac{1}{2}\int_{\partial V}\ud^{2}\vec{r}\;
\sigma_{d}(\vec{r})\Phi_{d}(\vec{r})
\end{equation}

We are dealing with a two-dimensional micrometric object, so the
surface energy term dominates over the volume energy term. We can
approximate this surface term by an effective easy plane
anisotropy contribution given by
\begin{equation}
 \mathcal{E}_{demag,S}=\int_{V}\ud^{3}\vec{r}\;2\pi M_{z}^{2}(\vec{r})
\end{equation}
This gives for the total energy
\begin{align}
\label{Energy}
 \mathcal{E}(\Phi,m)&=A\left[\frac{1}{1-m^2}(\nabla m)^2+(1-m^2)
 (\nabla\Phi)^2\right]\nonumber\\
 &-K_{\shortparallel}\cos^2\Phi(1-m^2)+(K_{\perp}+2\pi M_{s}^{2})m^2
\end{align}
and the equations of motion \eqref{LL2} become:
\begin{align}
\label{Eqmotion}
\frac{M_{s}}{\gamma}\frac{d\Phi}{dt}=&-\frac{2Am}{(1-m^2)^2}
(\nabla m)^2-\frac{2A}{1-m^2}\bigtriangleup m\nonumber\\
&- 2Am(\nabla\Phi)^2+2K_{\shortparallel}\cos^2\Phi\;m\nonumber\\
&+2(K_{\perp}+2\pi M_{s}^{2})m\nonumber\\
\frac{M_{s}}{\gamma}\frac{dm}{dt}=&-K_{\shortparallel}\sin(2\Phi)
(1-m^2)-4Am\nabla m\cdot\nabla\Phi\nonumber\\
&+2A(1-m^2)\bigtriangleup\Phi
\end{align}
In the slow dynamics regime ($|\dot{\vec{X}}|\ll1$) solutions
$(\Phi,m)$ can be split into $\Phi=\Phi_{0}+\Phi_{1}$ and
$m=m_{0}+m_{1}$, where $\Phi_{0}$ and $m_{0}$ are the static
solutions of the Landau-Lifshitz equation (we consider the
anisotropy interaction to be weak enough so that the static
solutions of the Hamiltonian
$\mathcal{E}_{ex}+\mathcal{E}_{demag}$ are valid for our problem)
and where $\Phi_{1}$ and $m_{1}$ are linear on $|\dot{\vec{X}}|$.
Static solutions are given by Eqs. \eqref{zeroth}. As discussed in
Section III, in the weak bending regime we can approximate
$\tilde{r}\simeq r$ so that $\ds
\nabla\Phi_{0}=n_v\frac{\hat{e}_{\phi}}{r}$ and $\ds \nabla
m_{0}=\frac{dm_{0}}{dr}\hat{e}_{r}$.

Linearizing Eqs. \eqref{Eqmotion} and taking into account that
$\ds \frac{d\Phi}{dt}=-\dot{\vec{X}}\cdot\nabla\Phi$ and that
$\ds \frac{dm}{dt}=-\dot{\vec{X}}\cdot\nabla m$, we obtain
the equations of motion
\begin{align}
 \label{Lineqsmotion2}
-\frac{M_{s}}{\gamma}&n_v\dot{\vec{X}}\cdot\frac{\hat{e}_{\phi}}{r}
=\frac{-2A}{1-m_{0}^{2}}\bigtriangleup m_{1}-\Bigg[\frac{2A}
{(1-m_{0}^{2})^2}\left(\frac{dm_{0}}{dr}\right)^2\nonumber\\
&+\frac{2A
}{r^2}-2K_{\shortparallel}\cos^{2}\Phi_{0}
-2(K_{\perp}+2\pi M_{s}^2)\Bigg]m_{1}\nonumber\\
&-\frac{4A\;m_{0}}{(1-m_{0}^{2})^2}\frac{dm_{0}}{dr}
\hat{e}_{r}\cdot\nabla m_{1}-4A n_v\;m_{0}\frac{\hat{e}_{\phi}}
{r}\cdot\nabla\Phi_{1}\nonumber\\
&-2K_{\shortparallel}m_{0}\sin(2\Phi_{0})
\Phi_{1}\nonumber\\
-\frac{M_{s}}{\gamma}&\dot{\vec{X}}\cdot\hat{e}_{r} \frac{dm_{0}}
{dr}=2A(1-m_{0}^{2})\bigtriangleup\Phi_{1}\nonumber\\
&-4An_v\;m_{0}\nabla m_{1}\cdot \frac{\hat{e}_{\phi}}{r}-4A\;m_{0}\frac{dm_{0}}{dr}\hat{e}_{r}\cdot\nabla\Phi_{1}\nonumber\\
&+2K_{\shortparallel}\sin(2\Phi_{0})m_{0}m_{1}-2K_{\shortparallel}\cos(2\Phi_{0})(1-m_{0}^2)\Phi_{1}
\end{align}

Asymptotic expressions for the $O(|\dot{\vec{X}}|)$ corrections to
the out-of-plane vortex shape can be determined by substituting
Eqs. \eqref{zeroth} into Eqs. \eqref{Lineqsmotion2}. In doing so
we obtain
\begin{align}
\label{solinfty}
m_{1}&=-\frac{M_{s}}{2\gamma}n_v\frac{\dot{\vec{X}}\cdot\hat{e}_{\phi}}
{(K_{\perp}+2\pi M_{s}^{2})+K_{\shortparallel}\cos^{2}\Phi_{0}}\frac{1}{r}\nonumber\\
\Phi_{1}&=\frac{C_{2}M_{s}}{2\gamma A}\Delta_{0}^{3/2}(\dot{\vec{X}}\cdot\hat{e}_{r})
\frac{\exp{(-r/\Delta_{0})}}{r^{1/2}}
\end{align}
for $r\gg\Delta_{0}$, and
\begin{align}
\label{solzero} m_{1}&=\frac{M_{s} C_{1}n_v}{3\gamma
A\Delta_{0}^{2}}(\dot{\vec{X}}
\cdot\hat{e}_{\phi})r^{3}\nonumber\\
\Phi_{1}&=\frac{M_{s}}{\gamma}\frac{p}{18A}
\dot{\vec{X}}\cdot\vec{r}
\end{align}
for $r\ll\Delta_{0}$. Computation of the mass of the vortex core
can be made via $\vec{\Pi}_{t}$, which should be proportional to
$\dot{\vec{X}}$ in this limit:
\begin{align}
 \label{momentum2}
 \vec{\Pi}_{t}&=-\frac{M_{s}}{\gamma}\int\ud^{2}\vec{r}(\nabla\Phi)m\nonumber\\
 &=-\frac{M_{s}}{\gamma}\int\ud^{2}\vec{r}(\nabla\Phi_{0})m_{0}-\frac{M_{s}}{\gamma}\int\ud^{2}\vec{r}(\nabla\Phi_{0})m_{1}\nonumber\\
&\quad-\frac{M_{s}}{\gamma}\int\ud^{2}\vec{r}(\nabla\Phi_{1})m_{0}-\frac{M_{s}}{\gamma}\int\ud^{2}\vec{r}(\nabla\Phi_{1})m_{1}
\end{align}
Notice that $\ds
-\frac{M_{s}}{\gamma}\int\ud^{2}\vec{r}(\nabla\Phi_{0})m_{0}=\vec{0}$
because it corresponds to the momentum of the static solution. The
last term of Eq. \eqref{momentum2} can be neglected because it is
quadratic in $|\dot{\vec{X}}|$. Therefore it remains to calculate
the second and third terms, which are given by
\begin{align}
&-\frac{M_{s}}{\gamma}\int\ud^{2}\vec{r}(\nabla\Phi_{0})m_{1}=-\frac{M_{s}}{\gamma}\int_{r\leq\Delta_{0}}\ud^{2}\vec{r}
 (\nabla\Phi_{0})m_{1}\nonumber\\
 &\qquad\qquad-\frac{M_{s}}{\gamma}\int_{r\geq\Delta_{0}}
 \ud^{2}\vec{r}(\nabla\Phi_{0})m_{1}=\\
 &
\frac{2\pi}{\gamma^{2}}\left(\frac{M_{s}^{2}}{K_{\perp}+2\pi
 M_{s}^{2}}\frac{1/4}{\sqrt{1+\frac{K_{\shortparallel}}
 {K_{\perp}+2\pi M_{s}^{2}}}}\ln(R/\Delta_{0})
-\frac{1}{56}\right)\dot{\vec{X}}\nonumber
\end{align}
and
\begin{align}
 -\frac{M_{s}}{\gamma}&\int\ud^{2}\vec{r}(\nabla\Phi_{1})m_{0}=
-\frac{M_{s}}{\gamma}\int_{r\leq\Delta_{0}}\ud^{2}\vec{r}
(\nabla\Phi_{1})m_{0}\nonumber\\
&\qquad\qquad-\frac{M_{s}}{\gamma}\int_{r\geq\Delta_{0}}
\ud^{2}\vec{r}(\nabla\Phi_{1})m_{0}\nonumber\\
&=\frac{2\pi}{\gamma^{2}}\left(-\frac{11}{504}
+\frac{2}{49}\left(1-\Xi\cdot e^2\right)\right)\dot{\vec{X}},
\end{align}
respectively. Notice that $\ds \Xi=\int_{1}^{R/\Delta_{0}} \ud x
\frac{\exp(-2x)}{x}\simeq\int_{1}^{\infty} \ud x
\frac{\exp(-2x)}{x}=0.049$ because we are interested in the limit $R\gg\Delta_{0}$.

Collecting all terms for the momentum, we get for the total mass density
\begin{align}
 \label{mass}
\rho_{M}=&\frac{2\pi}{\gamma^{2}}\left[
\frac{M_{s}^{2}}
{K_{\perp}+2\pi M_{s}^{2}}\frac{\ln(R/\Delta_{0})}{4\sqrt{1+
\frac{K_{\shortparallel}}{K_{\perp}+2\pi M_{s}^{2}}}}-0.0014\right]
\end{align}

Notice that we are interested in the limit $R\gg\Delta_{0}$, so
that the term involving $\ln(R/\Delta_{0})$ is the dominant one.
Furthermore, redefining the exchange length by a factor close to
unity we can always absorb the small numerical constant in Eq.\
(\ref{mass}) into the logarithmic term. Magneto-crystalline
anisotropies, if they are sufficiently large, destroy the
circularly polarized state. Consequently, materials like
permalloy, used in the studies of the vortex state, have
negligible magneto-crystalline anisotropy energy as compared to
the demagnetizing energy. This means that the above expression for
the vortex mass density can be reduced to
\begin{equation}
 \label{mass3}
\rho_{M}\simeq\frac{1}{4\gamma^{2}}\ln(R/\Delta_{0})
\end{equation}
With account of this formula one obtains the following expressions
for the parameters $\omega_{M}$ and $\epsilon_{0}$ that determine
eigenfrequencies in the equation (\ref{freq2}):
\begin{align}
\label{parameters}
\omega_{M}&=\frac{8\pi\gamma M_{s}}{\ln(R/\Delta_{0})}\nonumber\\
\epsilon_{0}&=\frac{5 L}{18\pi R}\ln(R/\Delta_{0})
\end{align}

\section{Effects of the magnetic field and dissipation}
In this section we study the effects of a magnetic field on the
excitation modes of the vortex state. Arbitrary directed magnetic
field can be split into two components, one being in the plane of
the disk and the other one being perpendicular to it. The effects
of these two components can be investigated separately.

Consider first the case of a spatially uniform in-plane magnetic
field, $\vec{H}_{in}=h_{x}\hat{e}_{x}+h_{y}\hat{e}_{y}$. For small
displacements along the disk, the magnetic vortex develops an
in-plane magnetization density given by\cite{Guslienko2}
\begin{equation}
\label{Magnetization}
\vec{M}(\vec{X})=-\mu\left[\hat{z}\times\vec{X}\right], \qquad
\mu=(2\pi/3)M_{s} n_v R.
\end{equation}
The Zeeman energy density term is
\begin{align}
\omega_{Z}(\vec{X})&=-M(\vec{X})\cdot\vec{H}_{in}=
-\mu\left[\hat{z}\times\vec{H}_{in}\right]\cdot\vec{X}\nonumber\\
&=\mu h_{y} x- \mu h_{x} y
\end{align}
and thus the total in-plane potential energy becomes
\begin{align}
\label{Potential2}
\omega_{XY}&(\vec{X})=\frac{1}{2}\rho_{M}\omega_{M}^2\epsilon_{0}\left(x^2+y^2\right)+\mu h_{y}x-\mu h_{x}y\nonumber\\
&=\frac{1}{2}\rho_{M}\omega_{M}^2\epsilon_{0}\Bigg[\left(x+\frac{\mu h_{y}}
{\rho_{M}\omega_{M}^2\epsilon_{0}}\right)^2\nonumber\\
&\quad+\left(y-\frac{\mu h_{x}}{\rho_{M}\omega_{M}^2\epsilon_{0}}\right)^2\Bigg]-\frac{1}{2}
\frac{\mu^2}{\rho_{M}\omega_{M}^2\epsilon_{0}}\vec{H}_{in}^2
\end{align}
Notice that by shifting the origin of the coordinate system we
retrieve the original in-plane term of the total energy density \eqref{Potential} except
for the constant term $\ds
-\frac{1}{2}\frac{\mu^2}{\rho_{M}\omega_{M}^2\epsilon_{0}}\vec{H}_{in}^2$,
which is field dependent. Consequently, the application of an
in-plane magnetic field does not modify the excitation modes given
by \eqref{freq2}.

Consider now the effect of the magnetic field perpendicular to the
plane of the disk. $\vec{H}_{\perp}=H\hat{z}$. Application of such
a field results in the precession of the magnetic moment of the
vortex about the direction of the field, described by the
Landau-Lifshitz equation\cite{CT-lectures},
\begin{equation}
\label{LL} \frac{\partial \vec{M}(t,\vec{X})}{\partial
t}=-\gamma\left[\vec{M}(t,\vec{X}) \times\vec{H}_{\perp}\right],
\end{equation}
where $\gamma$ is the electron gyromagnetic ratio. Formally, this
effect can be accounted for by adding an extra term to the
gyrovector. Indeed, integration of Eq. \eqref{Thiele} (with no
potential energy) on time gives
$\dot{\vec{X}}=\alpha\left[\vec{X}\times\vec{\rho}_{G}\right]$, where
$\alpha=-1/\rho_{M}$. With account of Eq. \eqref{Magnetization}, we have
\begin{align}
\left[\hat{z}\times\dot{\vec{X}}\right]&=-\gamma\left[\hat{z}
\times\vec{X}\right]\times\vec{H}_{\perp}\\
\alpha\left(\hat{z}\times\left[\vec{X}\times\vec{\rho}_{G}\right]\right)&=
-\gamma\left[\hat{z}\times\vec{X}\right]\times\vec{H}_{\perp}
\end{align}
The vector identity $\vec{a}\times\vec{b}\times\vec{c}=
(\vec{a}\cdot\vec{c})\vec{b}-(\vec{a}\cdot\vec{b})\vec{c}\;$
leads to $\alpha \rho_{G}=-\gamma H$. Consequently, the
precessional effect of the perpendicular field can be absorbed
into the gyrovector density if one adds to it the term $\ds
\vec{\rho}_{G,\vec{H}_{\perp}}=-\frac{\gamma}{\alpha}\vec{H}_{\perp}=\rho_{M}\gamma\vec{H}_{\perp}$.
This adds the Larmor frequency to $\omega_{M}$:
\begin{equation}
\omega_{M}(H)=\frac{\rho_{G,tot}}{\rho_{M}}=\omega_{M}+\frac{\rho_{G,\vec{H}_{\perp}}}
{\rho_{M}}=\omega_{M}+\gamma H
\end{equation}
so that the eigenfrequencies \eqref{freq2} become
\begin{equation}
\label{freq3}
\omega_{n}^{\pm}(H)=\frac{1}{2}\left[\sqrt{(1+4\epsilon(H))+\frac{4\lambda}{\rho_{M}\omega_{M}^{2}(H)}
q_{n}^{2}}\pm1\right]\omega_{M}(H)
\end{equation}
with $\epsilon(H)$ given by
\begin{equation}
\label{epsilon}
\epsilon(H)=\frac{\omega_{G}(H)}{\omega_{M}(H)}=\frac{\omega_{XY}^{''}(\vec{X}=\vec{0})} {\rho_{M}\omega_{M}^2(H)}=\frac{\epsilon_{0}}{(1+\gamma
H/\omega_{M})^2}
\end{equation}

Introducing dimensionless variables $h=\gamma H/\omega_{M}$ and
$\bar{\omega}_{n}^{\pm}(h)=\omega_{n}^{\pm}(H)/\omega_{M}$ we can rewrite
Eqs. \eqref{freq3} as
\begin{align}
\label{freq4}
\bar{\omega}_{n}^{\pm}(h)&=\frac{1}{2}\left[\sqrt{1+\frac{4\epsilon_{0}}{(1+h)^2}+\frac{4\lambda}{\rho_{M}\omega_{M}^{2}(1+h)^{2}}
q_{n}^{2}}\pm1\right]\times\nonumber\\
&\qquad(1+h)\nonumber\\
&\simeq\frac{1}{2}\left[\sqrt{1+\frac{4\epsilon_{0}}{(1+h)^2}}\pm1\right](1+h)\nonumber\\
&\qquad\qquad+\frac{\textrm{sgn}(1+h)}{\sqrt{(1+h)^2+4\epsilon_{0}}}\frac{\lambda q_{n}^{2}}{\rho_{M}\omega_{M}^{2}}
\end{align}
The distance between $\omega_{n}^{+}$ and $\omega_{n}^{-}$ equals $\Delta
\omega=\omega_{M}+\gamma H$.

To conclude this Section, we investigate the effects of the dissipation on
the excitation modes of magnetic vortices. We consider only the
zero field case. Derivation of the corresponding expressions when
a magnetic field is applied is straightforward. The way to
introduce dissipation into our equations is by adding a damping
term of the form $-D\dot{\vec{X}}$ ($D$ being the damping
constant) to Eq. \eqref{Thiele}\cite{Guslienko2,Thiele}.
Therefore, the elastic Thiele's equation becomes
\begin{equation}
\label{Damping}
\rho_{M}\ddot{\vec{X}}-\lambda\partial^{2}_{z}\vec{X}+\dot{\vec{X}}\times\vec{\rho}_{G}-D\dot{\vec{X}}+\rho_{M}\omega_{M}^2\epsilon_{0}\vec{X}=0
\end{equation}
Repeating the procedure of Sec. IV  with the above equation in the massive vortex case ($\rho_{M}\neq0$) we obtain the following equation for the frequency modes
\begin{equation}
\label{eqz}
\omega^{2}+(\omega_{M}+id)\omega-\omega_{M}^2\epsilon(q)=0
\end{equation}
with $d=D/\rho_{M}$ and $\epsilon(q)=\epsilon_{0}+\frac{\lambda}{\rho_{M}\omega_{M}^{2}}q^{2}$. The (complex) roots of this equation, $\omega_{\pm}=\textrm{Re}(\omega_{\pm})+i\textrm{Im}(\omega_{\pm})$, are given by
\begin{eqnarray}
\textrm{Re}(\omega_{\pm})&=&\mp\frac{r^{1/2}}{2}\cos(\theta/2)
-\frac{\omega_{M}}{2},\nonumber\\
\quad\textrm{Im}(\omega_{\pm})&=&\mp\frac{r^{1/2}}{2}\sin(\theta/2)-\frac{d}{2}
\end{eqnarray}
with
\begin{align}
\label{polar}
r&=\sqrt{\left[(1+4\epsilon(q))\omega_{M}^2-d^2\right]^2+4d^2\omega_{M}^2}\nonumber\\
\theta&=\arg{\Big(\left[(1+4\epsilon(q))\omega_{M}^2-d^2\right]+
i\left[2d\omega_{M}\right]\Big)}\nonumber\\
&=\arctan{\left(\frac{2d\omega_{M}}
{(1+4\epsilon(q))\omega_{M}^2-d^2}\right)}
\end{align}

In the regime of weak dissipation, $d<<\omega_{M}$, we
have $\ds
\theta\simeq\arctan\left[\frac{2d}{(1+4\epsilon(q))\omega_{M}}\right]$
and $r\simeq(1+4\epsilon(q))\omega_{M}^2$. As $\ds
\cos[\arctan(x)/2]\simeq1-\frac{x^2}{8}+o(x^4)$ and $\ds
\sin[\arctan(x)/2]\simeq\frac{x}{2}+o(x^3)$ if $|x|\ll1$, we
finally obtain
\begin{align}
\label{ReIm}
\textrm{Re}(\omega_{\pm})&=\mp\left[\frac{1}{2}\left(\sqrt{1+4\epsilon(q)}\pm1\right)
-\frac{1}{4}\frac{(d/\omega_{M})^2}{(1+4\epsilon(q))^{3/2}}\right]\omega_{M}\nonumber\\
&\simeq\mp\Bigg[\frac{\omega_{M}}{2}\left(\sqrt{1+4\epsilon_{0}}\pm1\right)
-\frac{\omega_{M}}{4}\frac{(d/\omega_{M})^2}{(1+4\epsilon_{0})^{3/2}}\nonumber\\
&+\frac{\lambda}{\sqrt{1+4\epsilon_{0}}}\left(1+\frac{3}{2}\frac{(d/\omega_{M})^{2}}{(1+4\epsilon_{0})^{2}}\right)\frac{q^{2}}{\rho_{M}\omega_{M}}\Bigg]
\end{align}
and
\begin{align}
\textrm{Im}(\omega_{\pm})&=\left(\mp\frac{1}{\sqrt{1+4\epsilon(q)}}-1\right)\frac{d}{2}\\
\frac{\textrm{Im}(\omega_{+})}
{\textrm{Im}(\omega_{-})}&=-\frac{1+\sqrt{1+4\epsilon(q)}}{1-
\sqrt{1+4\epsilon(q)}}=\frac{(1+\sqrt{1+4\epsilon(q)})^2}{4\epsilon(q)}
\end{align}

\section{Conclusions}
We have studied excitation modes of vortices in circularly
polarized mesoscopic magnetic disks that correspond to the
string-like gyroscopic waves in the vortex core. This problem was
studied by classical treatment based upon Landau-Lifshitz equation
and by quantum treatment based upon Hamiltonian approach. The
quantum problem is interesting on its own as it is equivalent to
the problem of quantum oscillations of a charged string confined
in a parabolic potential and subjected to the magnetic field,
which in its turn, is a generalization of the problem of the
field-induced orbital motion of the electron in a potential well.
Both treatments rendered identical results. Our solution
generalizes the expression for the frequency of the gyroscopic
motion of the vortex for the case of the finite wave number $q$,
as $\omega_-(q) = \omega_G + \gamma
M_s(q\Delta_0)^2\ln(R/\Delta_0)$, where $\omega_G$ is the
conventional gyrofrequency, $\gamma$ is the gyromagnetic ratio,
$M_s$ is the saturation magnetization, $\Delta_0$ is the exchange
length, and $R$ is the radius of the disk. This expression is
valid in the long-wave limit $q\Delta_0 \ll 1$. The wave number is
quantized, $q_n = 2\pi n/L$, where $L$ is the thickness of the
disk and $n$ is an integer. For a disk of radius $R \sim 1\mu$m,
thickness $L \sim 100$nm, exchange length $\Delta_0 \sim 5$nm, and
saturation magnetization $M_s \sim 10^3$emu, the $n = 1$ mode is
separated from $\omega_G$ by a few GHz. It could be excited by,
e.g., a tip of a force microscope or a micro-SQUID placed at the
center of the disk. Such measurement, while challenging, is
definitely within experimental reach.

Throughout this paper we considered of a non-zero mass of the
vortex. In addition to the gyroscopic mode the finite provides a
new excitation mode, $\omega_+(q) = \omega_M + \gamma
M_s(q\Delta_0)^2\ln(R/\Delta_0)$. The gap, $\omega_M$ is higher
than the gyroscopic frequency $\omega_G$. It depends explicitly on
the vortex mass. The vortex mass density has been computed by us
as a coefficient of proportionality, $\rho_{M}$, in the kinetic
energy of the moving vortex $\rho_{M}v^2/2$. It is given by
$\rho_{M}\simeq {1}/(4\gamma^{2})\ln(R/\Delta_{0})$, where $R$ is
the radius of the disk, $\Delta_0$ is the exchange length, and
$\gamma$ is the gyromagnetic ratio. For a $25$nm thick, micron
size permalloy disk this gives the vortex mass in the ball park of
$10^{-23}$kg, which is close to the experimental value estimated
for a comparable size permalloy ring\cite{Bedau}. Our result for
the mass gives $\omega_M = 8\pi \gamma M_{s}/\ln(R/\Delta_{0})$.
This is in the ballpark of, or below, the uniform ferromagnetic
resonance of the disk. It would be interesting to investigate this
frequency range experimentally alongside with the low-frequency
gyroscopic mode. One can also test in experiment the explicit
field dependence of the vortex modes, computed in this paper. So
far we have done it for the low field that only slightly disturbs
the vortex state formed in a zero field. However, the statement
concerning the existence of the additional mode due to the finite
vortex mass should apply to higher fields as well. This case,
however, defies analytical study and must employ full-scale
numerical micromagnetic calculations. When the field is sufficient
to fully polarize the disk in the perpendicular direction, we
expect the high frequency mode to evolve into the uniform
ferromagnetic resonance.

\section{Acknowledgements}
The work at the University of Barcelona was supported by the
Spanish Government Project No. MAT2008-04535. R.Z. acknowledges
financial support from the Ministerio de Ciencia e Innovaci\'{o}n
de Espa\~na. The work of E.M.C. at Lehman College is supported by
the Department of Energy through grant No. DE-FG02-93ER45487.

\end{document}